\documentclass[12pt,prd,tightenlines,nofootinbib,showpacs,showkeys]{revtex4}
\newcommand{\be}{\begin{equation}}
\newcommand{\ee}{\end{equation}}
\newcommand{\bdis}{\begin{displaymath}}
\newcommand{\edis}{\end{displaymath}}
\newcommand{\bga}{\begin{equation}\begin{gathered}}
\newcommand{\ega}{\end{gathered}\end{equation}}
\usepackage{bm}
\usepackage{graphics}
\usepackage{rotating}
\usepackage{epsfig}
\usepackage{amsmath}
\usepackage{amsfonts}

\begin{document}

\title{
\begin{flushright}
{\rm\normalsize SSU-HEP-06/12}
\end{flushright}
Lamb shift in muonic helium ion}
\author{\firstname{A.P.} \surname{Martynenko}}
\email{a.p.martynenko@samsu.ru}
\affiliation{Samara State University, 443011, Pavlov street 1, Samara, Russia}

\begin{abstract}
The Lamb shift $(2P_{1/2}-2S_{1/2})$ in the muonic helium ion
$(\mu~^4_2He)^+$ is calculated with the account of contributions of
orders $\alpha^3$, $\alpha^4$, $\alpha^5$ and $\alpha^6$. Special
attention is given to corrections of the electron vacuum
polarization, the nuclear structure and recoil effects. The obtained
numerical value of the Lamb shift $1379.028~meV$ can be considered as
a reliable estimate for the comparison with experimental data.
\end{abstract}

\pacs{31.30.Jv, 12.20.Ds, 32.10.Fn}

\keywords{muonic helium ion, Lamb shift}

\maketitle

\section{Introduction}

The ion of muonic helium $(\mu~^4_2He)^+$ is the bound state of the
negative muon and alpha particle. This simple atom is short-lived.
The lifetime is determined by the muon decay in a time
$\tau_\mu=2.19703(4)\cdot 10^{-6}$ s. The lepton mass increases when
passing from the electron to the muon hydrogenic atoms
($m_\mu/m_e=206.7682838(54)$ \cite{MT}). As a result the role of the
nuclear structure and polarizability effects, the electron vacuum
polarization corrections and the recoil contributions to the fine
and hyperfine structure of the energy spectrum essentially augments.
Muonic atoms represent a unique laboratory for the determination of
the nuclear properties. The effect of finite nuclear size is
particularly important for muonic atoms, in which the muonic wave
function has a significant overlap with the nucleus. Another nuclear
effect is nuclear polarizability which refers to the contribution
from intermediate excited states of the nucleus. So, the
experimental investigation of the $(2P-2S)$ Lamb shift in light
muonic atoms (muonic hydrogen, muonic deuterium, muonic helium ions)
can give more precise values of the nuclear charge radii (the
proton, deuteron, helion, alpha particle) \cite{EGS,SGK}. In the
case of muonic hydrogen the Lamb shift measurement is carried out at
present at PSI (Paul Sherrer Institute) \cite{PSI1,PSI2}. The
experiment in the field of laser spectroscopy of muonic helium ions
has been performed many years ago at the muon beam of the CERN
synchrocyclotron \cite{CERN,KJ}. The experiment has observed
resonances in the ion of muonic helium $(\mu~^4_2He)^+$ at $811.68
(15)$ nm and $897.6 (3)$ nm corresponding to the
$(2P_{3/2}-2S_{1/2})$ and $(2P_{1/2}-2S_{1/2})$ transitions. The
values of frequencies in meV are equal 1527.5 meV and 1381.29 (46)
meV. At a later time the experimental study of the $(2P-2S)$
splitting in the $(\mu~^4_2He)^+$ \cite{Hauser} found no resonance
effect at the wavelength interval $811.4\leq\lambda\leq 812.0$ nm
with a greater than $95\%$ probability. So, at present there is the
need of new experiment which could resolve the existing experimental
problem.

Theoretical investigation of the Lamb shift $(2P-2S)$ in muonic
helium ions was performed many years ago in
Refs.\cite{BR1,BR2,BR3,Drake} on the basis of the Dirac equation
(see other references in review article \cite{BR3}). Their
calculation took into account different QED corrections with the
accuracy 0.01 meV. High order corrections over the fine structure
constant $\alpha$ to the Lamb shift $(2P-2S)$ in the electron
hydrogenic atom were obtained in the last years in the analytical
form. Modern status of these calculations is presented in
Refs.\cite{EGS,SGK}. The aim of the present work is to calculate the
Lamb shift $(2P-2S)$ in the ion of muonic helium $(\mu~^4_2He)^+$
with the account of contributions of orders $\alpha^3$, $\alpha^4$,
$\alpha^5$ and $\alpha^6$ on the basis of quasipotential method in
quantum electrodynamics \cite{M1,M2,M3}. We consider such effects of
the electron vacuum polarization, the recoil and nuclear structure
corrections which are crucial to attain the high accuracy. With the
exception of the nuclear polarizability contribution, we calculate
all corrections in the interval $(2P_{1/2}-2S_{1/2})$ with a
precision 0.001 meV. Our purpose consists in the improvement of the
earlier performed calculations \cite{BR2,BR3} and derivation the
reliable estimate for the $(2P_{1/2}-2S_{1/2})$ Lamb shift, which
can be used for the comparison with experimental data. Modern
numerical values of fundamental physical constants are taken from
\cite{MT}: the electron mass $m_e=0.510998918(44)\cdot 10^{-3}$
GeV, the muon mass $m_\mu=0.1056583692(94)$ GeV, the fine structure
constant $\alpha^{-1}=137.03599911(46)$, the mass of alpha particle
$m_\alpha= 3.72737917(32)$ GeV.

\section{Effects of vacuum polarization in the one-photon interaction}

Our approach to the investigation of the Lamb shift $(2P-2S)$ in the
muonic helium ion $(\mu~^4_2He)^+$ is based on the use of quasipotential
method in quantum electrodynamics \cite{M3,M4,M5}, where the two-particle
bound state is described by the Schr\"odinger equation. The basic contribution
to the muon and $\alpha$-particle interaction operator is determined by
the Breit Hamiltonian \cite{t4}:
\begin{equation}
\label{eq:hb}
H_B=\frac{{\bf p}^2}{2\mu}-\frac{Z\alpha}{r}-\frac{{\bf p}^4}{8m_1^3}-
\frac{{\bf p}^4}{8m_2^3}+\frac{\pi Z\alpha}{2}\left(\frac{1}{m_1^2}+
\frac{1}{m_2^2}\right)\delta({\bf r})-
\end{equation}
\begin{displaymath}
-\frac{Z\alpha}{2m_1m_2r}\left({\bf p}^2+\frac{{\bf r}({\bf rp}){\bf p}}
{r^2}\right)+\frac{Z\alpha}{r^3}\left(\frac{1}{4m_1^2}+\frac{1}{2m_1m_2}\right)
({\bf L}{\mathstrut\bm\sigma}_1)=H_0+\Delta V^B,
\end{displaymath}
where $H_0={\bf p}^2/2\mu-Z\alpha/r$, $m_1$, $m_2$ are the muon and
$\alpha$-particle masses, $\mu=m_1m_2/(m_1+m_2)$.

The wave functions of $2S-$ and $2P$-states are equal:
\begin{equation}
\label{eq:psi}
\psi_{200}(r)=\frac{W^{3/2}}{2\sqrt{2\pi}}e^{-\frac{Wr}{2}}\left(1-\frac{Wr}{2}
\right),~~~\psi_{2lm}(r)=\frac{W^{3/2}}{2\sqrt{6}}e^{-\frac{Wr}{2}}WrY_{lm}(\theta,\phi),
~~W=\mu Z\alpha.
\end{equation}

\begin{figure}[htbp]
\centering
\includegraphics{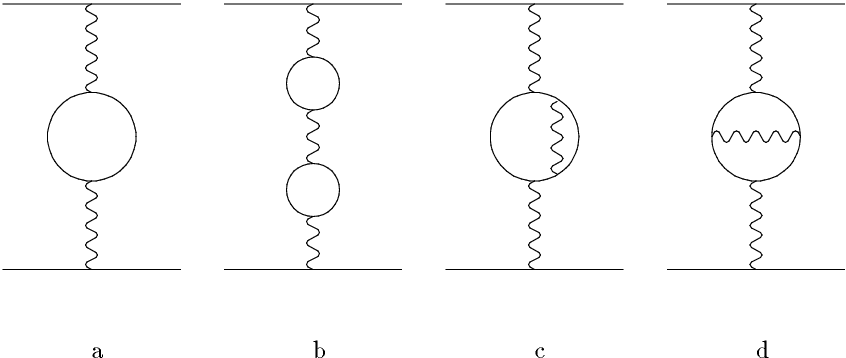}
\caption{Effects of one-loop and two-loop vacuum polarization in
the one-photon interaction.}
\label{fig:fig1}
\end{figure}

The ratio of the Bohr radius of muonic helium to the Compton wavelength of
the electron $m_e/W=0.34$, so, the basic contribution of the electron
vacuum polarization (VP) to the Lamb shift is of order $\alpha(Z\alpha)^2$
(see Fig.~\ref{fig:fig1}(a)).

Accounting the modification of the Coulomb potential due to vacuum
polarization in the coordinate representation
\begin{equation}
\label{eq:vcvp}
V^C_{VP}(r)=\frac{\alpha}{3\pi}\int_1^\infty d\xi \rho(\xi)
\left(-\frac{Z\alpha}{r}e^{-2m_e\xi r}\right),~~~
\rho(\xi)=\frac{\sqrt{\xi^2-1}(2\xi^2+1)}{\xi^4},
\end{equation}
we present one-loop VP contributions to the shifts of $2S-$,
$2P$-states and the Lamb shift $(2P-2S)$ in the form:
\begin{equation}
\label{eq:evp2s}
\Delta E_{VP}(2S)=-\frac{\mu(Z\alpha)^2\alpha}{6\pi}\int_1^\infty
\rho(\xi)d\xi\int_0^\infty x dx\left(1-\frac{x}{2}\right)^2e^{-x\left(1+
\frac{2m_e\xi}{W}\right)}=-2077.231~meV,
\end{equation}
\begin{equation}
\label{eq:evp2p}
\Delta E_{VP}(2P)=-\frac{\mu(Z\alpha)^2\alpha}{72\pi}\int_1^\infty
\rho(\xi)d\xi\int_0^\infty x^3 dx e^{-x\left(1+
\frac{2m_e\xi}{W}\right)}=-411.449~meV,
\end{equation}
\begin{equation}
\label{eq:evp2s2p}
\Delta E_{VP}(2P-2S)=1665.773~meV.
\end{equation}
The muon one-loop vacuum polarization correction is known in analytical form
\cite{EGS}. We included corresponding value to the total shift in section 5.
The two-loop vacuum polarization effects in the one-photon interaction are
shown in Fig.~\ref{fig:fig1}(b,c,d). To obtain the contribution of the amplitude in Fig.~\ref{fig:fig1}(b)
to the interaction operator, it is necessary to use the following replacement
in the photon propagator:
\begin{equation}
\frac{1}{k^2}\to\frac{\alpha}{3\pi}\int_1^\infty\rho(\xi)d\xi\frac{1}
{k^2+4m_e^2\xi^2}.
\end{equation}
In the coordinate representation the diagram with two sequential loops
gives the following particle interaction operator:
\begin{equation}
\label{eq:vvpvp}
V^C_{VP-VP}(r)=\frac{\alpha^2}{9\pi^2}
\int_1^\infty\rho(\xi)d\xi\int_1^\infty\rho(\eta)d\eta\left(-\frac{Z\alpha}{r}
\right)\frac{1}{(\xi^2-\eta^2)}\left(\xi^2e^{-2m_e\xi r}-\eta^2e^{-2m_e\eta r}
\right).
\end{equation}
Averaging (8) over the Coulomb wave functions (2), we find the contribution
to the Lamb shift of order $\alpha^2(Z\alpha)^2$:
\begin{equation}
\label{eq:evpvp}
\Delta E_{VP-VP}(2P-2S)=3.800~meV.
\end{equation}
Higher order $\alpha^2(Z\alpha)^4$ correction is determined by the amplitude
with two sequential electron (VP) and muon (MVP) loops. Corresponding
potential can be written as:
\begin{equation}
\label{eq:vvpmvp}
\Delta V_{VP-MVP}(r)=-\frac{4(Z\alpha)\alpha^2}{45\pi^2m_1^2}\int_1^\infty
\rho(\xi)d\xi\left[\pi\delta({\bf r})-\frac{m_e^2\xi^2}{r}e^{-2m_e\xi r}\right].
\end{equation}
Its contribution to the shift $(2P-2S)$ is equal
\begin{equation}
\label{eq:evpmvp}
\Delta E(2P-2S)=0.002~meV.
\end{equation}

\begin{figure}[htbp]
\centering
\includegraphics{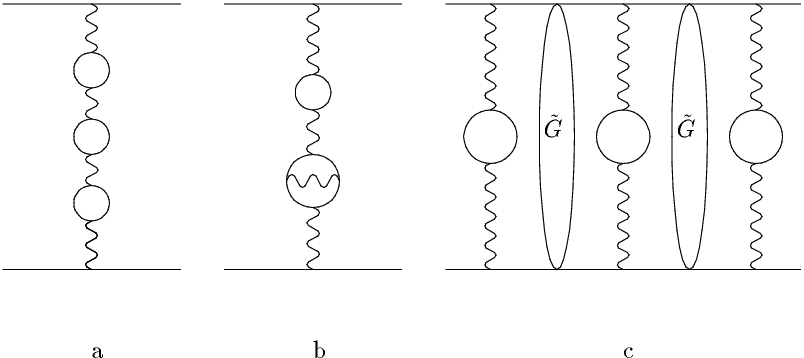}
\caption{Effects of the three-loop vacuum polarization in the
one-photon interaction (a,b) and in the third order perturbation
theory (c). $\tilde G$ is the reduced Coulomb Green function (33).}
\label{fig:fig2}
\end{figure}

The particle interaction potential, corresponding to two-loop amplitudes
in Fig.~\ref{fig:fig1}(c,d) with the second order polarization operator, takes the form:
\begin{equation}
\Delta V_{2-loop~VP}^C=-\frac{2}{3}\frac{Z\alpha}{r}
\left(\frac{\alpha}{\pi}\right)^2\int_0^1\frac{f(v)dv}{(1-v^2)}
e^{-\frac{2m_er}{\sqrt{1-v^2}}},
\end{equation}
where the spectral function
\begin{equation}
f(v)=v\Bigl\{(3-v^2)(1+v^2)\left[Li_2\left(-\frac{1-v}{1+v}\right)+2Li_2
\left(\frac{1-v}{1+v}\right)+\frac{3}{2}\ln\frac{1+v}{1-v}\ln\frac{1+v}{2}-
\ln\frac{1+v}{1-v}\ln v\right]
\end{equation}
\begin{displaymath}
+\left[\frac{11}{16}(3-v^2)(1+v^2)+\frac{v^4}{4}\right]\ln\frac{1+v}{1-v}+
\left[\frac{3}{2}v(3-v^2)\ln\frac{1-v^2}{4}-2v(3-v^2)\ln v\right]+
\frac{3}{8}v(5-3v^2)\Bigr\},
\end{displaymath}
$Li_2(z)$ is the Euler dilogarithm. The potential $\Delta
V^C_{2-loop~VP}(r)$ gives larger contribution as compared with (8)
both to the hyperfine structure and Lamb shift $(2P-2S)$:
\begin{equation}
\label{eq:e2loopvp}
\Delta E_{2-loop~VP}(2P-2S)=7.769~meV.
\end{equation}
Numerical value of corrections \eqref{eq:evpvp}, \eqref{eq:e2loopvp} and an accuracy of the
calculation show that it is important to consider three-loop
contributions of the vacuum polarization (see Fig.~\ref{fig:fig2}). A part of
corrections to the potential from the diagrams of three-loop vacuum
polarization in the one-photon interaction can be derived as the
relations (8), (12) (the sequential loops in Fig.~\ref{fig:fig2}(a,b)).
Corresponding contributions to the potential and the splitting
$(2P-2S)$ are the following:
\begin{equation}
V^C_{VP-VP-VP}(r)=-\frac{Z\alpha}{r}\frac{\alpha^3}{(3\pi)^3}\int_1^\infty
\rho(\xi)d\xi\int_1^\infty\rho(\eta d\eta\int_1^\infty\rho(\zeta)d\zeta\times
\end{equation}
\begin{displaymath}
\times\left[e^{-2m_e\zeta r}\frac{\zeta^4}{(\xi^2-\zeta^2)(\eta^2-\zeta^2)}
+e^{-2m_e\xi r}\frac{\xi^4}{(\zeta^2-\xi^2)(\eta^2-\xi^2)}+
e^{-2m_e\eta r}\frac{\eta^4}{(\xi^2-\eta^2)(\zeta^2-\eta^2)}\right],
\end{displaymath}
\begin{equation}
V^C_{VP-2-loop~VP}=-\frac{4\mu\alpha^3(Z\alpha)}{9\pi^3}\int_1^\infty
\rho(\xi)d\xi\int_1^\infty\frac{f(\eta)d\eta}{\eta}\frac{1}{r(\eta^2-\xi^2)}
\left(\eta^2e^{-2m_e\eta r}-\xi^2e^{-2m_e\xi r}\right),
\end{equation}
\begin{equation}
\Delta E_{VP-VP-VP}(2P-2S)=0.008~meV,
\end{equation}
\begin{equation}
\Delta E_{VP-2-loop~VP}(2P-2S)=0.036~meV.
\end{equation}

There exists a number of the diagrams that express three-loop
corrections in the polarization operator. They were first calculated
for the $(2P-2S)$ Lamb shift in Refs.\cite{KN1,KN2}. The largest
contribution to the energy spectrum comes from the sixth-order
vacuum polarization diagrams with one electron loop ($\Pi^{(p6)}$
corrections \cite{KN1}). The estimate of their contribution to the
Lamb shift in $(\mu~^4_2He)^+$ is included in Table~\ref{tb1}. The analysis
of the contribution of three-loop vacuum polarization in the third
order perturbation theory in Fig.~\ref{fig:fig2}(c) shows that we can neglect it
accounting the declared accuracy of the calculation.

\begin{figure}[htbp]
\centering
\includegraphics{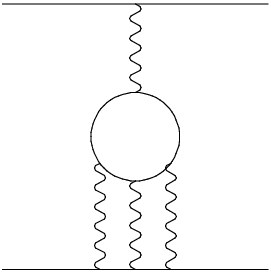}
\caption{The Wichmann-Kroll correction. The wave line shows the Coulomb
photon.}
\label{fig:fig3}
\end{figure}

Additional one-loop vacuum polarization diagram is presented in Fig.~\ref{fig:fig3}.
In the energy spectrum it gives the correction of the fifth
order over $\alpha$ (the Wichmann-Kroll correction) \cite{WK,MPS}.
The particle interaction potential can be written in this case in
the integral form:
\begin{equation}
\Delta V^{WK}(r)=\frac{\alpha(Z\alpha)^3}{\pi r}\int_0^\infty \frac{d\zeta}
{\zeta^4} e^{-2m_e\zeta r}\left[-\frac{\pi^2}{12}\sqrt{\zeta^2-1}\theta(\zeta-1)
+\int_0^\zeta dx\sqrt{\zeta^2-x^2} f^{WK}(x)\right].
\end{equation}
The exact form of the spectral function $f^{WK}$ is presented in
Refs.\cite{EGS,WK,MPS}. Numerical integration in Eq.(19) with the wave
functions (2) gives the following contribution to the Lamb shift:
\begin{equation}
\Delta E^{WK}(2P-2S)=-0.020~meV.
\end{equation}

\section{Relativistic corrections with the vacuum polarization effects}

The electron vacuum polarization effects lead not only to corrections in
the Coulomb potential \eqref{eq:vcvp}, but also to the modification of the other terms
of the Breit Hamiltonian \eqref{eq:hb}. The one-loop vacuum polarization corrections
in the Breit interaction were obtained in Refs.\cite{KP1,KP2}:
\begin{equation}
\Delta V^B_{VP}(r)=\frac{\alpha}{3\pi}\int_1^\infty\rho(\xi)d\xi\sum_{i=1}^4
\Delta V_{i,VP}^B(r),
\end{equation}
\begin{equation}
\Delta V_{1,VP}^B=\frac{Z\alpha}{8}\left(\frac{1}{m_1^2}+\frac{1}{m_2^2}\right)
\left[4\pi\delta({\bf r})-\frac{4m_e^2\xi^2}{r}e^{-2m_e\xi r}\right],
\end{equation}
\begin{equation}
\Delta V_{2,VP}^B=-\frac{Z\alpha m_e^2\xi^2}{m_1m_2r}e^{-2m_e\xi
r}(1- m_e\xi r),
\end{equation}
\begin{equation}
\Delta V_{3,VP}^B=-\frac{Z\alpha}{2m_1m_2}p_i\frac{e^{-2m_e\xi r}}{r}
\left[\delta_{ij}+\frac{r_ir_j}{r^2}(1+2m_e\xi r)\right]p_j,
\end{equation}
\begin{equation}
\Delta V_{4,VP}^B=\frac{Z\alpha}{r^3}\left(\frac{1}{4m_1^2}+\frac{1}{2m_1m_2}
\right)e^{-2m_e\xi r}(1+2m_e\xi r)({\bf L}{\mathstrut\bm\sigma}_1).
\end{equation}

In the first order perturbation theory (PT) the potentials
$\Delta V_{i,VP}^B(r)$ give necessary contributions of order
$\alpha(Z\alpha)^4$ to the shift $(2P-2S)$:
\begin{equation}
\Delta E_{1,VP}^B(2P-2S)=-0.894~meV,
\end{equation}
\begin{equation}
\Delta E_{2,VP}^B(2P-2S)=0.012~meV,
\end{equation}
\begin{equation}
\Delta E_{3,VP}^B(2P-2S)=0.022~meV,
\end{equation}
\begin{equation}
\Delta E_{4,VP}^B(2P-2S)=-0.088~meV.
\end{equation}
The potentials $\Delta V_{2,VP}^B$, $\Delta V_{3,VP}^B$, $\Delta
V_{4,VP}^B$ take into account the recoil effects over the ratio
$m_1/m_2$. We have included in Table I the summary correction of
order $\alpha(Z\alpha)^4$, which is determined by the relations
(26)-(29). Next to leading order correction of order
$\alpha^2(Z\alpha)^4$ appears in the energy spectrum from
two-loop modification of the Breit Hamiltonian. We consider the term
of the leading order over $m_1/m_2$ in the potential (the function
$f(v)$ is determined by Eq.(13)):
\begin{equation}
\Delta V_{2-loop~VP}^B(r)=\frac{\alpha^2(Z\alpha)}{12\pi^2}\left(\frac{1}
{m_1^2}+\frac{1}{m_2^2}\right)\int_0^1\frac{f(v)dv}{1-v^2}\left[4\pi
\delta({\bf
r})-\frac{4m_e^2}{(1-v^2)r}e^{-\frac{2m_er}{\sqrt{1-v^2}}}\right].
\end{equation}
Corresponding $(2P-2S)$ shift is the following:
\begin{equation}
\Delta E_{2-loop~VP}^B(2P-2S)=-0.003~meV.
\end{equation}
Other two-loop contributions to the Breit potential are omitted because
they give the energy corrections which lie outside the accuracy of the
calculation in this work.

\begin{figure}[htbp]
\centering
\includegraphics{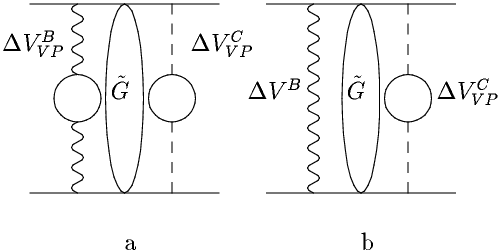}
\includegraphics{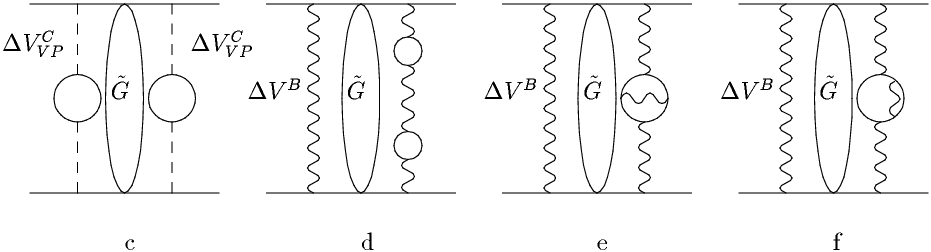}
\caption{Effects of the one-loop and two-loop vacuum polarization in
the second order perturbation theory (SOPT). The dashed line shows
the Coulomb photon. $\tilde G$ is the reduced Coulomb Green function
(33). The potentials $\Delta V^B$, $\Delta V^C_{VP}$ and $\Delta
V^B_{VP}$ are determined respectively by relations (1), (3) and
(21).}
\label{fig:fig4}
\end{figure}

In the second order perturbation theory (SOPT) we have a number of
the electron vacuum polarization contributions in the leading orders
$\alpha^2(Z\alpha)^2$ and $\alpha(Z\alpha)^4$, shown in the diagrams
of Fig.4 (b,c):
\begin{equation}
\Delta E_{SOPT}^{VP}=<\psi|\Delta V^C_{VP}\tilde G\Delta V^C_{VP}|\psi>+
2<\psi|\Delta V^B\tilde G\Delta V^C_{VP}|\psi>
\end{equation}
The second order perturbation theory corrections in the energy spectrum of
hydrogen-like system are determined by the reduced Coulomb Green function
$\tilde G$ (RCGF), whose partial expansion has the form \cite{VP}:
\begin{equation}
\tilde G_n({\bf r}, {\bf r'})=\sum_{l,m}\tilde g_{nl}(r,r')Y_{lm}({\bf n})
Y_{lm}^\ast({\bf n'}).
\end{equation}
The radial function $\tilde g_{nl}(r,r')$ was presented in \cite{VP} in
the form of the Sturm expansion in the Laguerre polynomials. For the calculation
of the Lamb shift $(2P-2S)$ in muonic helium it is convenient to use
the compact representation for the RCGF of $2S-$ and $2P-$ states,
which was obtained in \cite{KP1}:
\begin{equation}
\tilde G(2S)=-\frac{Z\alpha\mu^2}{4x_1x_2}e^{-\frac{x_1+x_2}{2}}\frac{1}
{4\pi}g_{2S}(x_1,x_2),
\end{equation}
\begin{equation}
g_{2S}(x_1,x_2)=8x_<-4x^2_<+8x_>+12x_<x_>-26x^2_<x_>+2x^3_<x_>-4x^2_>-
26x_<x^2_>+23x^2_<x^2_>-
\end{equation}
\begin{displaymath}
-x^3_<x^2_>+2x_<x^3_>-x^2_<x^3_>+4e^x(1-x_<)(x_>-2)x_>+4(x_<-2)x_<(x_>-2)x_>
\times
\end{displaymath}
\begin{displaymath}
\times[-2C+Ei(x_<)-\ln(x_<)-\ln(x_>)],
\end{displaymath}
\begin{equation}
\tilde G(2P)=-\frac{Z\alpha\mu^2}{36x^2_1x^2_2}e^{-\frac{x_1+x_2}{2}}\frac{3}
{4\pi}\frac{({\bf x}_1{\bf x}_2)}{x_1x_2}g_{2P}(x_1,x_2),
\end{equation}
\begin{equation}
g_{2P}(x_1,x_2)=24x^3_<+36x^3_<x_>+36x^3_<x^2_>+24x^3_>+36x_<x^3_>+36x^2_<
x^3_>+49x^3_<x^3_>-3x^4_<x^3_>-
\end{equation}
\begin{displaymath}
-12e^x_<(2+x_<+x_<^2)x^3_>-3x^3_<x^4_>+12x_<^3x_>^3[-2C+Ei(x_<)-\ln(x_<)-
\ln(x_>)],
\end{displaymath}
where $x_<=min(x_1,x_2)$, $x_>=max(x_1,x_2)$, $C=0.57721566...$ is
the Euler constant. As a result the two-loop vacuum polarization
contributions in the first term of Eq.(32) can be presented
originally in the integral form (Fig.4(c)). The subsequent numerical
integration gives the following results:
\begin{equation}
\Delta E^{VP,VP}_{SOPT}(2S)=-\frac{\mu\alpha^2(Z\alpha)^2}{72\pi^2}
\int_1^\infty\rho(\xi)d\xi\int_1^\infty\rho(\eta)d\eta\times
\end{equation}
\begin{displaymath}
\times\int_0^\infty\left(1-\frac{x}{2}\right)e^{-x\left(1-\frac{2m_e\xi}{W}
\right)}dx\int_0^\infty\left(1-\frac{x'}{2}\right)
e^{-x'\left(1-\frac{2m_e\eta}{W}\right)}dx'g_{2S}(x,x')=-1.901~meV,
\end{displaymath}
\begin{equation}
\Delta E^{VP,VP}_{SOPT}(2P)=-\frac{\mu\alpha^2(Z\alpha)^2}{7776\pi^2}\int_1^\infty
\rho(\xi)d\xi\int_1^\infty\rho(\eta)d\eta\times
\end{equation}
\begin{displaymath}
\times\int_0^\infty e^{-x\left(1+\frac{2m_e\xi}{W}\right)}dx
\int_0^\infty e^{-x'\left(1+\frac{2m_e\eta}{W}\right)}
dx'g_{2P}(x,x')=-0.194~meV,
\end{displaymath}

The second term in (32) has the similar structure (see Fig.4(b)). The
transformation of the different matrix elements is carried out
with the use of the algebraic relation of the form:
\begin{equation}
<\psi|\frac{{\bf p}^4}{(2\mu)^2}{\sum}'_m\frac{|\psi_m><\psi_m|}{E_2-E_m}
\Delta V^C_{VP}|\psi>=<\psi|(E_2+\frac{Z\alpha}{r})(\hat H_0+
\frac{Z\alpha}{r}){\sum}'_m\frac{|\psi_m><\psi_m|}{E_2-E_m}\Delta
V_{VP}^C |\psi>=
\end{equation}
\begin{displaymath}
=<\psi|\left(E_2+\frac{Z\alpha}{r}\right)^2\tilde G\Delta V_{VP}^C|\psi>-
<\psi|\frac{Z\alpha}{r}\Delta V_{VP}^C|\psi>+<\psi|\frac{Z\alpha}{r}|\psi>
<\psi|\Delta V_{VP}^C|\psi>.
\end{displaymath}
Omitting further details of the calculation of numerous matrix
elements in Eq.(40), we present here the summary numerical
contribution from the second term in Eq.(32) to the shift $(2P-2S)$:
\begin{equation}
\Delta E^{B,VP}_{SOPT}(2P-2S)=1.434~meV.
\end{equation}

\begin{figure}[htbp]
\centering
\includegraphics{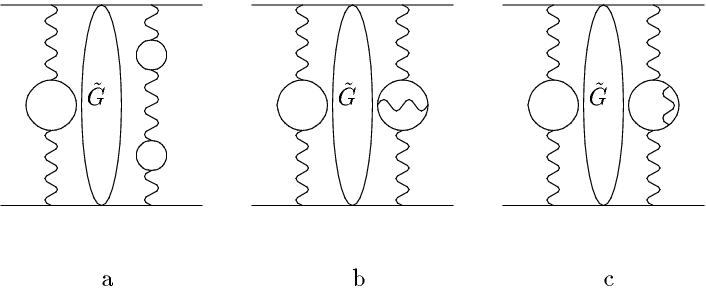}
\caption{The three-loop vacuum polarization corrections in the
second order perturbation theory. $\tilde G$ is the reduced Coulomb
Green function.}
\label{fig:fig5}
\end{figure}

The three-loop vacuum polarization contributions to the energy spectrum
in the second order perturbation theory are presented in Fig.5.
Respective potentials are obtained earlier in relations (3), (8), (12).
Considering the accuracy of the calculation we can restrict our
analysis by the shifts of $2S-$ level, which can be written in the form:
\begin{equation}
\Delta E^{VP-VP,VP}_{SOPT}(2S)=-\frac{\mu\alpha^3(Z\alpha)^2}{108\pi^3}
\int_1^\infty\rho(\xi)d\xi\int_1^\infty\rho(\eta)d\eta\int_1^\infty
\rho(\zeta)d\zeta\int_0^\infty dx(1-\frac{x}{2})\times
\end{equation}
\begin{displaymath}
\int_0^\infty dx'(1-\frac{x'}{2})
e^{-x'(1+\frac{2m_e\zeta}{W})}\frac{1}{\xi^2-\eta^2}\left[\xi^2
e^{-x(1+\frac{2m_e\xi}{W})}-\eta^2e^{-x(1+\frac{2m_e\eta}{W})}\right]g_{2S}
(x,x')=-0.011~meV,
\end{displaymath}
\begin{equation}
\Delta E^{2-loop~VP,VP}_{SOPT}(2S)=-\frac{\mu\alpha^3(Z\alpha)^2}{18\pi^3}
\int_0^1\frac{f(v)dv}{1-v^2}\int_1^\infty\rho(\xi)d\xi\times
\end{equation}
\begin{displaymath}
\times \int_0^\infty
dx\left(1-\frac{x}{2}\right)e^{-x(1+\frac{2m_e}{\sqrt{1-v^2}W})}
\int_0^\infty
dx'\left(1-\frac{x'}{2}\right)e^{-x'(1+\frac{2m_e\xi}{W})}g_{2S}(x,x')
=-0.017~meV,
\end{displaymath}
Yet another contributions of the second order PT exist (see
Fig.4(d,e,f)), which have the general structure similar to Eqs.(38),
(39). They appear after the replacements $\Delta V_{VP}^C\to \Delta
V^B$ and $\Delta V^C_{VP}\to \Delta V^C_{VP,VP}$ in the basic
amplitude shown in Fig.4(c). The estimate of this contribution of
order $\alpha^2(Z\alpha)^4$ to the shift $(2P-2S)$ can be derived if
we take into account in the Breit potential the leading order term
in the ratio $m_1/m_2$. Its numerical value is
\begin{equation}
\Delta E_{SOPT}^{VP,VP;\Delta V^B}(2P-2S)=0.008~meV.
\end{equation}
The two-loop vacuum polarization contribution is determined also by
the amplitude in Fig.4(a). To obtain its numerical value in the
energy spectrum we have to use Eqs.(3) and (21). In the leading
order in the ratio $m_1/m_2$ we take again the potential (22), which
leads to the following correction of order $\alpha^2(Z\alpha)^4$:
\begin{equation}
\Delta E_{SOPT}^{VP,\Delta V^B_{VP}}(2P-2S)=-0.006~meV.
\end{equation}

\section{Nuclear structure and vacuum polarization effects}

\begin{figure}[htbp]
\centering
\includegraphics{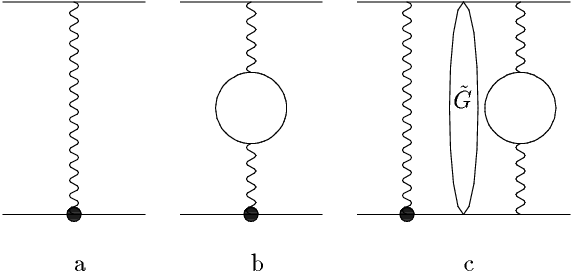}
\caption{The leading order nuclear structure and vacuum polarization
corrections. The thick point represents the nuclear vertex
operator.}
\label{fig:fig6}
\end{figure}

The influence of the nuclear structure on the muon motion in the ion
$(\mu~^4_2He)^+$ is determined in the leading order by the charge radius
of alpha particle $r_\alpha=1.676(8)~fm$ \cite{Friar,Sick} (Fig.6(a)):
\begin{equation}
\Delta E_{str}(2P-2S)=-\frac{\mu^3(Z\alpha)^4}{12}<r^2_\alpha>=-295.848~meV.
\end{equation}

Next to leading order correction of order $(Z\alpha)^5$ is described
by one-loop exchange diagrams (Fig.7). Introducing the charge form
factor $F(k^2)$ of the alpha particle, we can express it in the integral
form:
\begin{equation}
\Delta E_{str}^{2\gamma}(nS)=-\frac{\mu^3(Z\alpha)^5}{\pi n^3}\delta_{l0}
\int_0^\infty\frac{dk}{k}V(k),
\end{equation}
\begin{equation}
V(k)=\frac{2(F^2-1)}{m_1m_2}+\frac{8m_1[-F(0)+4m_2^2F'(0)]}{m_2(m_1+m_2)k}
+\frac{k^2}{2m_1^3m_2^3}\times
\end{equation}
\begin{displaymath}
\times\left[2(F^2-1)(m_1^2+m_2^2)-F^2m_1^2\right]
+\frac{\sqrt{k^2+4m_1^2}}{2m_1^3m_2(m_1^2-m_2^2)k}\times
\end{displaymath}
\begin{displaymath}
\times\Biggl\{k^2\left[2(F^2-1)m_2^2-F^2m_1^2\right]
+8m_1^4F^2+\frac{16m_1^4m_2^2(F^2-1)}{k^2}\Biggr\}-
\end{displaymath}
\begin{displaymath}
-\frac{\sqrt{k^2+4m_2^2}m_1}{2m_2^3(m_1^2-m_2^2)k}\Biggl\{k^2
\left[2(F^2-1)-F^2\right]+8m_2^2F^2+\frac{16m_2^4(F^2-1)}{k^2}\Biggr\}.
\end{displaymath}
To perform numerical integration in Eq.(47) we use the dipole and Gaussian
parameterizations of the charge form factor:
\begin{equation}
F(k^2)=\frac{\Lambda^4}{(k^2+\Lambda^2)^2},~\Lambda^2=\frac{12}{<r^2_\alpha>},~
F(k^2)=\exp[-\frac{1}{6}k^2r^2_\alpha].
\end{equation}
Numerical values of contributions to the Lamb shift $(2P-2S)$ are equal
\begin{equation}
\Delta E_{D,str}^{2\gamma}(2P-2S)=7.196~meV,~\Delta E_{G,str}^{2\gamma}(2P-2S)=6.605~meV.
\end{equation}

\begin{figure}[htbp]
\centering
\includegraphics{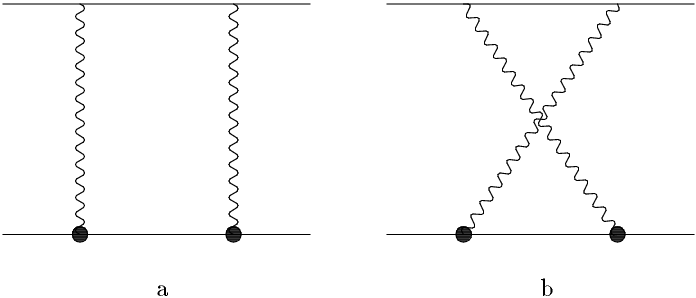}
\caption{The nuclear structure corrections of order $(Z\alpha)^5$.
The thick point is the nuclear vertex operator.}
\label{fig:fig7}
\end{figure}

The essential increase of corrections (46), (50) as compared with
muonic hydrogen is conditioned by two reasons. In the first place,
the charge radius of the $\alpha$ - particle increases so that
$\mu^2\cdot <r_\alpha^2>=0.76$. Secondly, the charge of the
$\alpha$-particle $Z=2$, resulting the additional factor $2^5$ in
Eq.(50). The particle interaction amplitudes containing the nuclear
structure and vacuum polarization effects must be considered to
obtain total value of the Lamb shift. In the leading order such
amplitude is presented in Fig.6(b). Corresponding interaction
operator can be written as
\begin{equation}
\Delta V^{VP}_{str}(r)=\frac{2}{3}\pi Z\alpha<r^2_\alpha>\frac{\alpha}{3\pi}
\int_1^\infty\rho(\xi)d\xi\left[\delta({\bf r})-\frac{m_e^2\xi^2}{\pi r}
e^{-2m_e\xi r}\right].
\end{equation}
Its contributions to the shifts of the $2S-$ and $2P-$ levels are determined
by the following expressions:
\begin{equation}
\Delta E_{str}^{VP}(2S)=\frac{\alpha(Z\alpha)^4<r_\alpha^2>\mu^3}{36\pi}
\int_1^\infty\rho(\xi)d\xi[1-\frac{4m_e^2\xi^2}{W^2}\int_0^\infty
x dx (1-\frac{x}{2})^2e^{-x(1+\frac{2m_e\xi}{W})}]=0.937~meV,
\end{equation}
\begin{equation}
\Delta E^{VP}_{str}(2P)=-\frac{\alpha(Z\alpha)^4\mu^3<r_\alpha^2>}{108\pi}
\frac{m_e^2}{W^2}\int_1^\infty \xi^2\rho(\xi)d\xi\int_0^\infty
x^3e^{-x(1+\frac{2m_e\xi}{W})}dx=-0.023~meV,
\end{equation}
\begin{equation}
\Delta E^{VP}_{str}(2P-2S)=-0.960~meV.
\end{equation}

The contribution of the same order $\alpha(Z\alpha)^4$ is given by the
amplitude in the second order perturbation theory in Fig.6(c):
\begin{equation}
\Delta E^{VP}_{str,SOPT}(2P-2S)=-\frac{\alpha(Z\alpha)^4\mu^3<r_\alpha^2>}
{36\pi}\int_1^\infty\rho(\xi)d\xi\times
\end{equation}
\begin{displaymath}
\times\int_0^\infty dx e^{-x(1+\frac{2m_e\xi}{W})}(1-\frac{x}{2})
\left[4x(x-2)(\ln x+C)+x^3-13x^2+6x+4\right]=-1.506~meV.
\end{displaymath}

\begin{figure}[htbp]
\centering
\includegraphics{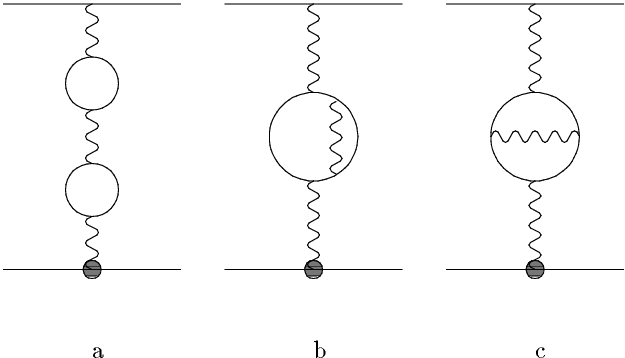}
\caption{The nuclear structure and two-loop vacuum polarization
effects in the one-photon interaction. The thick point is the
nuclear vertex operator.}
\label{fig:fig8}
\end{figure}

The two-loop vacuum polarization corrections with the account of the
nuclear structure are presented in Fig.8(a,b,c). The interaction potentials
constructed by means of Eqs.(7), (8), (12), (51), are determined by the
integral relations:
\begin{equation}
\Delta V^{VP-VP}_{str}(r)=\frac{2}{3}Z\alpha<r_\alpha^2>\left(\frac{\alpha}
{3\pi}\right)^2\int_1^\infty\rho(\xi)d\xi\int_1^\infty\rho(\eta)d\eta\times
\end{equation}
\begin{displaymath}
\times\left[\pi\delta({\bf r})-\frac{m_e^2}{r(\xi^2-\eta^2)}\left
(\xi^4 e^{-2m_e\xi r}-\eta^4e^{-2m_e\eta r}\right)\right],
\end{displaymath}
\begin{equation}
\Delta V^{2-loop~VP}_{str}(r)=\frac{4}{9}Z\alpha<r_\alpha^2>\left
(\frac{\alpha}{\pi}\right)^2\int_0^1\frac{f(v)dv}{1-v^2}\left[\pi\delta({\bf r})-
\frac{m_e^2}{r(1-v^2)}e^{-\frac{2m_er}{\sqrt{1-v^2}}}\right].
\end{equation}
The sum of corrections (56) and (57) to the Lamb shift $(2P-2S)$ is equal:
\begin{equation}
\Delta E_{str}^{VP,VP}(2P-2S)=-0.008~meV.
\end{equation}

\begin{figure}[htbp]
\centering
\includegraphics{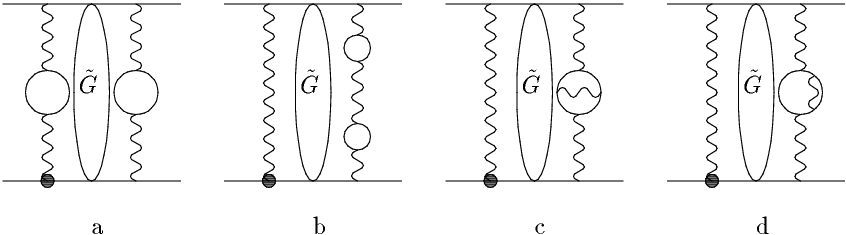}
\caption{The nuclear structure and two-loop vacuum polarization
effects in the second order perturbation theory. The sick point is
the nuclear vertex operator. $\tilde G$ is the reduced Coulomb Green
function.}
\label{fig:fig9}
\end{figure}

We have included in the Table I the two-loop vacuum polarization
and nuclear structure contribution of order $\alpha^2(Z\alpha)^4$
$\Delta E_{str,SOPT}^{VP,VP}$ in the second order PT, shown in Fig.9(a,b,c,d).
In the sixth order over $\alpha$ there exists also the nuclear structure
correction coming from the two-photon exchange diagrams with the electron vacuum
polarization insertion (see Fig.10). We find the analytical expression
of this correction and the numerical value by means of Eq.(47) in the form:
\begin{equation}
\Delta E^{2\gamma}_{str,VP}(nS)=-\frac{2\mu^3\alpha(
Z\alpha)^5}{\pi^2 n^3} \int_0^\infty k V(k) dk
\int_0^1\frac{v^2(1-\frac{v^2}{3})dv}{k^2(1-v^2)+4m_e^2},
\end{equation}
\begin{equation}
\Delta E_{str,VP}^{2\gamma}(2P-2S)=0.127~meV.
\end{equation}

\begin{figure}[htbp]
\centering
\includegraphics{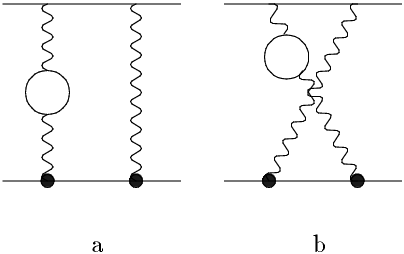}
\caption{The nuclear structure and electron vacuum polarization
effects in the two-photon exchange diagrams. The thick point is the
nuclear vertex operator.}
\label{fig:fig10}
\end{figure}

\section{Recoil corrections, muon self-energy and vacuum polarization
effects}

The investigation of the different order corrections to the Lamb
shift $(2P-2S)$ of electronic hydrogen has been performed for many
years. Modern analysis of the advances in the solution of this
problem is presented in a review articles \cite{EGS,SGK}. The most
part of the results was obtained in the analytical form, so they can
be used directly in the muonic helium ion. In this section we
analyse different contributions up to the sixth order over $\alpha$
in the energy spectrum $(\mu~^4_2He)^+$ and derive their numerical
values in the Lamb shift $(2P-2S)$.

The recoil correction of order $(Z\alpha)^4$ in the Lamb shift
appears in the matrix element of the Breit potential with the
functions (2) \cite{SY,EGS,PK}:
\begin{equation}
\Delta E_{rec}(2P-2S)=\frac{\mu^3(Z\alpha)^4}{12 m_2^2}=0.295~meV.
\end{equation}
The recoil correction of the fifth order over $(Z\alpha)$ is determined
by the relation \cite{SY,EGS}:
\begin{equation}
\Delta E_{rec}^{(Z\alpha)^5}=\frac{\mu^3(Z\alpha)^5}{m_1m_2\pi
n^3}\Bigl[\frac{2}{3}\delta_{l0}
\ln\frac{1}{Z\alpha}-\frac{8}{3}\ln
k_0(n,l)-\frac{1}{9}\delta_{l0}-
\frac{7}{3}a_n-\frac{2}{m_2^2-m_1^2}\delta_{l0}(m_2^2\ln\frac{m_1}{\mu}-
m_1^2\ln\frac{m_2}{\mu})\Bigr],
\end{equation}
where $\ln k_0(n,l)$ is the Bethe logarithm:
\begin{equation}
\ln k_0(2S)=2.811769893120563,
\end{equation}
\begin{equation}
\ln k_0(2P)=-0.030016708630213,
\end{equation}
\begin{equation}
a_n=-2\left[\ln\frac{2}{n}+(1+\frac{1}{2}+...+\frac{1}{n}+1-\frac{1}{2n}\right]\delta_{l0}+
\frac{(1-\delta_{l0})}{l(l+1)(2l+1)}.
\end{equation}
The expression (62) gives the following numerical result:
\begin{equation}
\Delta E_{rec}^{(Z\alpha)^5}(2P-2S)=-0.433~meV.
\end{equation}
The recoil correction of the sixth order over $(Z\alpha)$ was
calculated analytically in \cite{EG}:
\begin{equation}
\Delta E_{rec}^{(Z\alpha)^6}(2P-2S)=\frac{(Z\alpha)^6m_1^2}{8m_2}
\left(\frac{23}{6}-4\ln 2\right)=0.004~meV.
\end{equation}
The energy contributions obtained from the radiative corrections in
the lepton line, the Dirac and Pauli form factors and the muon
vacuum polarization have the form \cite{EG1,EGS}:
\begin{equation}
\Delta E_{MVP,MSE}(2S)=\frac{\alpha(Z\alpha)^4}{8\pi}\frac{\mu^3}{m_1^2}
\Biggl[\frac{4}{3}\ln\frac{m_1}{\mu(Z\alpha)^2}-\frac{4}{3}\ln k_0(2S)+
\frac{38}{45}+
\end{equation}
\begin{displaymath}
+\frac{\alpha}{\pi}\left(-\frac{9}{4}\zeta(3)+\frac{3}{2}
\pi^2\ln 2-\frac{10}{27}\pi^2-\frac{2179}{648}\right)+4\pi Z\alpha\left(
\frac{427}{384}-\frac{\ln 2}{2}\right)\Biggr]=10.939~meV,
\end{displaymath}
\begin{equation}
\Delta E_{MVP,MSE}(2P)=\frac{\alpha(Z\alpha)^4}{8\pi}\frac{\mu^3}{m_1^2}
\Biggl[-\frac{4}{3}\ln k_0(2P)-\frac{m_1}{6\mu}-
\end{equation}
\begin{displaymath}
-\frac{\alpha}{3\pi}\frac{m_1}{\mu}\left(\frac{3}{4}\zeta(3)-\frac{
\pi^2}{2}\ln 2+\frac{\pi^2}{12}+\frac{197}{144}\right)\Biggr]
=-0.168~meV.
\end{displaymath}
Omitting explicit form of the radiative-recoil corrections of orders
$\alpha(Z\alpha)^5$ and $(Z^2\alpha)(Z\alpha)^4$ from the Table 9
\cite{EGS}, we give their numerical value in the Lamb shift
$(2P-2S)$ of muonic helium $(\mu~^4_2He)^+$:
\begin{equation}
\Delta E_{rad-rec}(2P-2S)=-0.038~meV.
\end{equation}
The nuclear structure corrections of orders $(Z\alpha)^6$ and
$\alpha(Z\alpha)^5$ were studied in Refs.\cite{Friar1,LYE} for arbitrary
hydrogenic atom. Let us present their numerical values in the case of
muonic helium:
\begin{equation}
\Delta E_{str}^{(Z\alpha)^6}(2P-2S)=\frac{(Z\alpha)^6}{12}\mu^3\Bigl\{r_\alpha^2
\left[\langle\ln \mu Z\alpha r\rangle+C-\frac{3}{2}\right]-\frac{1}{2}r_\alpha^2+
\frac{1}{3}\langle r^3\rangle\langle\frac{1}{r}\rangle-
\end{equation}
\begin{displaymath}
-I_2^{rel}-I_3^{rel}
-\mu^2F_{NR}+\frac{1}{40}\mu^2\langle r^4\rangle\Bigr\}
=-0.12576\cdot r_\alpha^2+0.047=-0.306 ~meV,
\end{displaymath}
where the quantities $I_{2,3}^{rel}, F_{NR}$ are written explicitly in \cite{Friar1},
\begin{equation}
\Delta E_{str}^{\alpha(Z\alpha)^5}(2P-2S)=0.070~meV.
\end{equation}

\begin{figure}[htbp]
\centering
\includegraphics{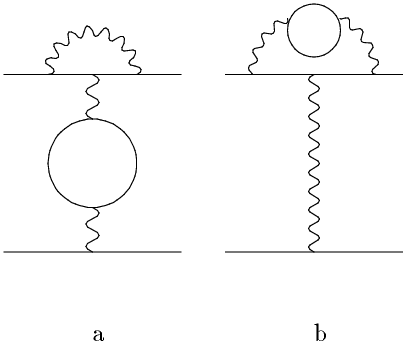}
\caption{Radiative corrections with the vacuum polarization effects.}
\label{fig:fig11}
\end{figure}

The diagram in Fig.~\ref{fig:fig11}(b) gives the contribution to the energy spectrum,
which can be expressed in terms of the slope of the Dirac form factor
$F_1'$ and the Pauli form factor $F_2$:
\begin{equation}
\Delta E_{rad+VP}(nS)=\frac{\mu^3}{m_1^2}\frac{(Z\alpha)^4}{n^3}\left[4m_1^2
F_1'(0)\delta_{l0}+F_2(0)\frac{C_{jl}}{2l+1}\right],
\end{equation}
\begin{equation}
C_{jl}=\delta_{l0}+(1-\delta_{l0})\frac{j(j+1)-l(l+1)-\frac{3}{4}}{l(l+1)}.
\end{equation}
The two-loop contribution to the form factors $F_1'(0)$ and $F_2(0)$ was
calculated in \cite{BCR}:
\begin{equation}
m_1^2F_1'(0)=\left(\frac{\alpha}{\pi}\right)^2\left[\frac{1}{9}\ln^2\frac
{m_1}{m_e}-\frac{29}{108}\ln\frac{m_1}{m_e}+\frac{1}{9}\zeta(2)+\frac{395}
{1296}\right],
\end{equation}
\begin{equation}
F_2(0)=\left(\frac{\alpha}{\pi}\right)^2\left[\frac{1}{3}\ln\frac{m_1}{m_e}
-\frac{25}{36}+\frac{\pi^2}{4}\frac{m_e}{m_1}-4\frac{m_e^2}{m_1^2}\ln\frac{m_1}
{m_e}+3\frac{m_e^2}{m_1^2}\right].
\end{equation}
Then the correction to the Lamb shift $(2P-2S)$ is equal
\begin{equation}
\Delta E_{rad+VP}(2P-2S)=-0.031~meV.
\end{equation}
To estimate the muon self-energy and electron vacuum polarization
contribution in Fig.~\ref{fig:fig11}(a), we use the relation obtained in \cite{KP1}:
\begin{equation}
\Delta E^{VP}_{MSE}=\frac{\alpha}{3\pi
m_1^2}\ln\frac{m_1}{\mu(Z\alpha)^2} \left[<\psi_n|\Delta\cdot
\Delta V^C_{VP}|\psi_n>+2<\psi_n|\Delta V^C_{VP}\tilde G
\Delta\left(-\frac{Z\alpha}{r}\right)|\psi_n>\right].
\end{equation}
The sum of all matrix elements which appear in Eq.(80) leads to the
following shift $(2P-2S)$:
\begin{equation}
\Delta E^{VP}_{MSE}(2P-2S)=-0.107~meV.
\end{equation}
The hadron vacuum polarization (HVP) contribution can be taken into account
on the basis of numerical result obtained for muonic hydrogen
in Refs.\cite{borie-1981,Friar2,M6}. The HVP correction and the contribution of the nuclear
polarizability calculated in Refs.\cite{Rinker,JB} are included in Table I.

\section{Summary and conclusion}

In this work, various corrections of orders $\alpha^3$, $\alpha^4$,
$\alpha^5$ and $\alpha^6$ have been calculated for the Lamb shift
$(2P-2S)$ in muonic helium ion $(\mu~^4_2He)^+$. Contrary to earlier
performed investigations of the energy spectra of light muonic atoms
in Refs.\cite{BR1,BR2}, we have used the three-dimensional
quasipotential approach for the description of two-particle bound
state. Our analysis of the different contributions to the Lamb shift
accounts for the terms of two groups. The first group contains the
specific corrections for muonic helium ion, connected with the
electron vacuum polarization effects, nuclear structure and recoil
effects in the first and second order perturbation theory. The
contributions of this group are calculated numerically for the first
time. The necessary order corrections of the second group include
the analytical results known from the corresponding calculation in
the electronic hydrogen Lamb shift. Recent advances in the physics
of the energy spectra of simple atoms are presented in the review
articles \cite{EGS,SGK} which we use in this study. Numerical values
of all corrections are written in the Table I, which contains also
basic references on the earlier performed investigations (other
references can be found in \cite{EGS}). Total numerical value
1379.028 meV of the Lamb shift $(2P-2S)$ in muonic helium ion from
the Table I is in the agreement with theoretical results
$1380.9$ meV obtained in Refs.\cite{BR1,BR2} and 1378.71 meV \cite{borie2012},
and with the CERN
experimental value $1381.29$ meV \cite{CERN}. The difference of our
results from Refs.\cite{BR1,BR2} is connected both to the
calculation of new contributions of higher order and slightly
different numerical value of the charge radius of $\alpha$-particle
$r_\alpha$ used in this work. The authors of Refs.\cite{BR1,BR2}
used the value of charge radius $r_\alpha=1.674(12)~fm$. As has been
mentioned above the correction values were obtained with a 0.001 meV
accuracy because certain contributions to the Lamb shift $(2P-2S)$
of order $\alpha^6$ attain the value of several $\mu eV$. The
theoretical error caused by the uncertainties in the fundamental
parameters (fine structure constant, particle masses) entering the
Eq.(6) is around $10^{-5}$ meV. Other part of the theoretical
error is related to QED corrections of higher order. This part can
be estimated from the leading contribution of higher order over
$\alpha$: $m_1\alpha(Z\alpha)^6\ln(Z\alpha)/\pi n^3\approx 0.001$
meV. Finally, the biggest theoretical uncertainty connected with the
nuclear structure and polarizability contributions is discussed
below.

Let us summarize the basic particularities of the calculation performed
above.

1. Numerical value of specific parameter $m_e/\mu Z\alpha=0.34$
in muonic helium $(\mu~^4_2He)^+$ is sufficiently large, so the
electron vacuum polarization effects play essential role in the
interaction operator. We have considered one-loop, two-loop and
three-loop vacuum polarization contributions.

2. Nuclear structure effects are expressed in the Lamb shift of
muonic helium ion both in terms of the charge radius of $\alpha$ -
particle in the leading and next to leading orders and by means of
the charge form factor of the $\alpha$-particle in two-photon
exchange amplitudes.

3. An estimate of the alpha particle polarizability contribution to
the Lamb shift is taken from Refs.\cite{JB,Rinker}. Nuclear
structure and polarizability effects give the largest theoretical
uncertainty in the total value of the Lamb shift $(2P-2S)$. For
instance, in the leading order $(Z\alpha)^4$ the theoretical error,
connected with the uncertainty in the value of the alpha particle
charge radius $r_\alpha=1.676(8)~fm$ comprises $\pm 2.8~meV$.
Corresponding theoretical error of the nuclear polarizability
contribution amounts to 0.6 meV \cite{JB,Rinker}. Total
numerical result of this work for the $(2P-2S)$ Lamb shift can be
used for a comparison with the future experimental data and
determination more precise value of the alpha particle charge
radius.

\begin{acknowledgments}
The author is grateful to E.~Borie, R.N.~Faustov, F.~Kottmann and A.A.~Krutov for useful discussions
and critical remarks.
This work was supported by the Russian Foundation for Basic Research
(grant No. 06-02-16821, 14-02-00173).
\end{acknowledgments}

\begin{table}[htbp]
\caption{\label{t1}Lamb shift $(2P_{1/2}-2S_{1/2})$ in muonic
helium ion $(\mu~^4_2He)^+$.}
\bigskip
\label{tb1}
\begin{ruledtabular}
\begin{tabular}{|c|c|c|}  \hline
Contribution to the splitting &$\Delta E(2P-2S)$,~meV  & Formula, Reference   \\   \hline
1&2&3 \\  \hline
VP contribution of order $\alpha(Z\alpha)^2$ & 1665.773 & (6)  \\
in one-photon interaction &  &   \\   \hline
Two-loop VP contribution of order &11.569    &(9), (14)    \\
$\alpha^2(Z\alpha)^2$ in $1\gamma$ interaction  &   &   \\    \hline
VP and MVP contribution   &0.002     & (11)        \\
in $1\gamma$ interaction &     &     \\    \hline
Three-loop VP contribution   &0.046    &(17), (18), \cite{KN1}     \\
in $1\gamma$  interaction &    &     \\    \hline
The Wichmann-Kroll correction  & -0.020  &  (20)  \\  \hline
Additional LBL correction  & 0.006  &  \cite{borie2012,sgk2010}  \\  \hline
Relativistic and VP corrections of order  & -0.948    &  (26)-(29)    \\
$\alpha(Z\alpha)^4$ in the first order PT      &       &    \\    \hline
Relativistic and two-loop VP  & -0.003    &  (31)    \\
corrections of order $\alpha^2(Z\alpha)^4$   &       &    \\
in the first order PT   &       &    \\    \hline
Two-loop VP contribution of order   &1.707   &(38)-(39)     \\
$\alpha^2(Z\alpha)^2$ in the second order PT &    &    \\    \hline
Relativistic and one-loop VP  & 1.434    &  (41)    \\
corrections of order $\alpha(Z\alpha)^4$  &       &    \\
in the second order PT  &    &    \\    \hline
Three-loop VP contribution in the           &0.028  &(42)-(43)  \\
second order PT of order $\alpha^3(Z\alpha)^2$  &      &    \\    \hline
Relativistic and two-loop VP   & 0.002    &  (44)-(45)    \\
corrections of order $\alpha^2(Z\alpha)^4$  &       &    \\
in the second order PT  &    &    \\    \hline
Nuclear structure contribution of order $(Z\alpha)^4$   &-295.848  & (46),\cite{EGS} \\   \hline
Nuclear structure contribution   &6.605 (7.196)  & (50), \cite{borie2012}   \\
of order $(Z\alpha)^5$ from $2\gamma$ amplitudes                           &          &    \\   \hline
Nuclear structure and VP contribution  &-0.960  & (54)   \\
in $1\gamma$ interaction of order $\alpha(Z\alpha)^4$ &   &  \\   \hline
Nuclear structure and VP contribution   &-1.506  & (55)   \\
in the second order PT of order $\alpha(Z\alpha)^4$ &   &  \\   \hline
Nuclear structure and two-loop VP  &-0.008  & (58)   \\
contribution in $1\gamma$ interaction of order $\alpha^2(Z\alpha)^4$ &   &  \\   \hline
Nuclear structure and two-loop VP contribution&-0.018  & Fig.9   \\
in the second order PT of order $\alpha^2(Z\alpha)^4$ &   &  \\   \hline
Nuclear structure contribution of order   &0.127  & (60)   \\
$\alpha(Z\alpha)^5$ from $2\gamma$ amplitudes with VP insertion &          &    \\   \hline
\end{tabular}
\end{ruledtabular}
\end{table}
\begin{table}[htbp]
Table I (continued).\\
\bigskip
\begin{ruledtabular}
\begin{tabular}{|c|c|c|}  \hline
1&2&3 \\  \hline
Recoil correction of order $(Z\alpha)^4$   &0.295   &(61),\cite{SY,EGS,PK}   \\   \hline
Recoil correction of order $(Z\alpha)^5$   &-0.433   &(66),\cite{SY,EGS}   \\   \hline
Recoil correction of order $(Z\alpha)^6$   &0.004   &(67),\cite{EGS}   \\   \hline
Muon self-energy and MVP contribution &-11.107 & (68)-(69),\cite{EGS}  \\  \hline
Radiative-recoil corrections  &-0.038 &(70), Table 9 \cite{EGS}   \\
of orders $\alpha(Z\alpha)^5$, $(Z^2\alpha)(Z\alpha)^4$  &   &   \\  \hline
Nuclear structure corrections  & -0.236 &(71), (72) \cite{Friar1,LYE,EGS}  \\
of orders $(Z\alpha)^6$, $\alpha(Z\alpha)^5$ &  &  \\  \hline
Muon form factor $F_1'(0)$, $F_2(0)$ contributions & -0.031& (79),\cite{EGS,KP1,BCR}   \\  \hline
Muon self-energy and VP contribution &-0.107 (-0.065)& (81),\cite{KP1,EGS,Wundt}  \\  \hline
HVP contribution & 0.223 & \cite{borie-1981,Friar2,M6} \\  \hline
Nuclear polarizability contribution &2.470   &\cite{JB,Rinker,ji}  \\   \hline
Total contribution  & 1379.028  &  \\  \hline
\end{tabular}
\end{ruledtabular}
\end{table}

\end{document}